
\magnification=\magstephalf
\abovedisplayskip=5pt plus 3pt minus 1pt
\belowdisplayskip=5pt plus 3pt minus 1pt
\belowdisplayshortskip=5pt plus 3pt minus 2pt
\hsize=5.75truein
\vsize=8.75truein
\baselineskip=11pt
\nopagenumbers
\centerline{A BURST OF SPECULATION}
\bigskip
\centerline{J. I. Katz}
\centerline{Washington University, St. Louis, Mo. 63130}
\bigskip
\centerline{ABSTRACT}
\bigskip
Self-consistent models of gamma-ray burst source regions at 100 Kpc distance
are possible if the radiating plasma is confined to very thin sheets, and I
estimate parameters.  Energy sources
might be elastic (by starquakes) or magnetic (by reconnection), but
mechanisms remain obscure.  I discuss a very speculative model involving
collisions between comets in a hypothetical inner Oort cloud.
\bigskip
\centerline{INTRODUCTION}
\bigskip
Discoveries by the BATSE$^{1,2}$ have reduced theories (and theorists) of
gamma-ray bursts (GRB) to a state of confusion.  The field is nearly as open
as it was when GRB were first announced in 1973.  In order to understand
GRB, it will be necessary to answer some questions
about the physical conditions within them:
\item{1.} Is the energy of a GRB (or a sub-burst, in the case of GRB with
distinct substructure) released promptly in the form of energetic particles
and radiation, which gradually dissipate, or is the release continuous,
with an instantaneous power proportional to the instantaneously observed
intensity?
\item{2.} How do GRB time histories indicate characteristic physical time
scales?
\item{3.} Are there any intermediate reservoirs of energy between its
ultimate source and the radiating plasma (for example, vibrations in models
driven by elastic energy)?
\item{4.} Is the geometry of the source region line-like, sheet-like, or
sphere-like?
\item{5.} What is the optical depth of the source region?
\item{6.} Does an individual GRB have preferred power levels and decay
time scales, as may be suggested by repetitive sub-burst structure in at
least one observed GRB$^2$?
\item{7.} How isotropic or beamed is the radiation pattern?
\item{8.} Does the time history of observed intensity indicate the time
history of total radiated power, or is an anisotropic beam pattern variable
or rotationally modulated?
\item{9.} Is the continuum radiation mechanism annihilation, bremsstrahlung,
curvature radiation, or some other process?
\item{10.} Is the energy distribution of the particles in the source region
strongly nonthermal or partly thermalized?
\item{11.} Is the radiating plasma freely expanding or magnetically trapped?

The answers to these physical questions are closely related to the unsolved
astronomical questions of the distances to GRB and the existence of
counterparts in other energy bands.
The astronomical questions having so far proved intractable, the attention of
the theorist should turn to the physical questions.
\bigskip
\centerline{ESTIMATES}
\bigskip
Arguments$^{3,4}$ concerning the reported$^5$ 511 KeV annihilation line in
the March 5, 1979 GRB led to the inference that the emitting region was a
geometrically and optically thin sheet.  Although this soft gamma repeater
(SGR) may not be representative of GRB, and the status of the annihilation
line in GRB spectra is controversial, it may still be useful to consider
this geometry.  Magnetic reconnection, in analogy to a Solar flare, was one
of the first suggested models of GRB$^{4,6}$, and would be expected to
release energy in a thin sheet.

Consider a pair or electron-proton sheet plasma of thickness $L$, with particle
energies $\sim mc^2$, where $m$ is the electron mass, and density $n$.  The
transverse optical depth $\tau$ is
$$\tau \sim n \sigma_0 L, \eqno(1)$$
where the cross-section $\sigma_0 = (e^2/mc^2)^2$ is roughly applicable to
Coulomb and Compton scattering and to two photon annihilation and pair
production. The characteristic power density (per unit area) is
$$P \sim n^2 m c^3 \sigma_0 L \sim {n^2 e^4 L \over mc}. \eqno(2)$$
Define a dimensionless length $\ell \equiv L mc^2/e^2$ and power density
$p \equiv P \hbar^3/m^4c^6$.  Then
$$n \sim {\tau \over \ell \alpha^3}{m^3 c^3 \over \hbar^3}, \eqno(3)$$
where $\alpha \equiv e^2/\hbar c$ is the usual fine structure constant.  Use
of (1) tacitly assumes that $\tau$ is not large, as implied by the observed
nonthermal spectra.  Similarly, from (2):
$$n \sim \left({p \over \ell \alpha^3}\right)^{1/2} {m^3 c^3 \over \hbar^3}.
\eqno(4)$$
Equating (3) and (4) yields
$$\ell \sim {\tau^2 \over p \alpha^3}. \eqno(5)$$
These estimates lead to an energy content per unit area
$$\Sigma \sim n m c^2 L \sim \tau {m^3 c^6 \over e^4} \sim 1.0 \times
10^{19} \tau\ {\rm erg/cm}^2, \eqno(6)$$
and a radiation time scale (also the annihilation time scale $t_a \sim 1/n
\sigma_0 c)$
$$t_r \equiv {\Sigma \over P} \sim {\tau \over p}{\hbar^3 \over me^4} \sim 2
\times 10^{-17} \left({\tau \over p}\right)\ {\rm sec}. \eqno(7)$$

In order to maintain a radiating plasma there must be a continuing injection of
energy from an electric field.  An elementary estimate of the electrical
conductivity is
$$\sigma_{El} \sim {e \over m c \sigma_0} \sim {m c^3 \over e^2} \sim
10^{23}\ {\rm sec}^{-1}. \eqno(8)$$
The required electric field $E$ may be estimated from the condition of energy
balance
$$P = \sigma_{El} E^2 L, \eqno(9)$$
with the result
$$E \sim \left({p \alpha^3 \over \ell}\right)^{1/2} {m^2 c^4 \over e^3} \sim
{p \alpha^3 \over \tau}{m^2 c^4 \over e^3} \sim 2 \times 10^9 {p \over \tau}
\ {\rm cgs}. \eqno(10)$$

It is necessary to replenish annihilating pairs.  The
fastest processes, with cross-sections $\sim \sigma_0$, are $e^+ + e^- \to
\gamma + \gamma$ and $\gamma + \gamma \to e^+ + e^-$, which conserve the sum
of the numbers of leptons and photons.  To supply an escaping flux of
gamma-rays requires injection of both energy and particles.  The electric
field directly supplies only energy.  Particle number may be resupplied by
processes such as $\gamma + e^\pm \to e^\pm + \gamma + \gamma$, $e^+ + e^-
\to e^+ + e^- + \gamma$, $e^+ +e^- \to \gamma + \gamma + \gamma$, and
$\gamma + \gamma \to e^+ + e^- + \gamma$, with cross-sections $\sim \sigma_0
\alpha$.  Perhaps more important may be synchrotron radiation and magnetic
one-photon pair production, which occur because leptons and photons are
produced and scattered with large cross-field momenta.

The electric field (10) also produces a mean leptonic drift velocity $\sim
c$.  This may lead to a two-stream or ion-acoustic plasma instability,
constrained by the magnetic field to be one-dimensional.  The effective
conductivity may be reduced far below (8).  Such an increased
resistivity in regions of high current density is a familiar feature of
magnetic reconnection.  Quantitative understanding would require plasma
simulations which include collisional as well as collective processes.
\bigskip
\centerline{GAMMA-RAY BURSTS AT 100 KPC?}
\bigskip
The BATSE data$^1$ demonstrated the impossibility of a Galactic disc origin
of all GRB, and increased the attractiveness of earlier suggestions$^{7-9}$
that they are distributed in an extended halo of radius $\sim$ 100 Kpc.  If
so, then their luminosities approach those of the March 5, 1979 event in the
LMC, and many arguments$^{3-5,10,11}$ made for it may be applied to GRB in
general.

A typical GRB luminosity at 100 Kpc is $10^{41}$ erg/sec, corresponding to
an observed flux $\sim 10^{-7}$ erg/cm$^2$sec.  With an effective radiating
area of $10^{12}$ cm$^2$, $P \sim 10^{29}$ erg/cm$^2$sec and $p \sim 2
\times 10^{-7}$.  Then (5) yields $\ell \sim 10^{13} \tau^2$ and $L \sim 3
\tau^2$ cm.  In the absence of plasma instabilities $E \sim 500$ cgs,
insufficient to sustain a curvature radiation cascade.
It is evident that $t_r$ is extremely short unless $p$ is so small that the
GRB are within the Oort cloud.

A power density $P \sim 10^{29}$ erg/cm$^2$sec corresponds to a black body
of effective temperature $T_e \sim 20$ KeV, inconsistent with
the observed hard spectra.  At semi-relativistic energies
equilibration by Compton scattering is rapid, even in the absence of true
absorption, so that the more energetic part of the spectrum must be produced
in optically thin regions.  This argument need not apply
at photon energies $h\nu \ll 100$ KeV; self-absorption may limit reradiation
from a neutron star's surface.

At least one GRB reported at this meeting$^2$ apparently consisted of
distinct but very similar sub-bursts, each with a pronounced
time-skewness$^{12}$ consisting of a rapid rise and more gradual decay over
roughly one second.  This suggested injection of energy into a reservoir,
followed by its gradual radiation.  It is apparent that pair plasma cannot
be such a reservoir.  Although a magnetic field of $10^{12}$ gauss
provides sufficient stress to confine $10^{41}$ erg ($10^{20}$ gm) of pair
plasma, its opacity $\kappa \sim e^4 / m^3 c^4 \sim 100$ cm$^2$/gm would
lead to an optical depth $\tau \sim 10^{10}$, inconsistent with
the emergent hard and nonequilibrium spectrum.  If confined, the energy
would leak out over a time $\sim \tau r /c \sim 3 \times 10^5$ sec, also
inconsistent.  Similarly, its energy density of $\sim 10^{23}$ erg/cm$^3$
corresponds to a black-body temperature of $\sim 160$ KeV.  An unconfined
fireball of this temperature would lead to a flash of gamma-rays with mean
energy $\sim 500$ KeV, but with a thermal spectrum and a duration $\le r/3c
\sim 10^{-5}$ sec$^{10}$.  Thus pair gas cannot be an energy reservoir,
whether magnetically confined or exploding in a fireball; energy must be
continuously replenished.

An unconfined fireball continuously resupplied with energy will have a lower
energy density.  The energy density is $\sim P/c$, and $\tau \sim P r \kappa
/ 3 c^3 \sim 10^5$, still enough to ensure thermalization of the spectrum.
This problem is exacerbated at cosmological distances.  Adiabatic expansion
has the effect of collimating the particle and photon momenta, but
preserves the equilibrium spectral shapes, inconsistent with observation.

Models of the type discussed here suffer from the well-known problem of
a high pair-production optical depth for MeV gamma-rays.  Electric fields
produce opposing streams of $e^+$ and $e^-$ whose gamma-rays interact
head-on.  The well-known solution of relativistic collimation may work if
the radiating particles have sufficient energy and the majority are
collimated.  This is naturally obtained in an electrically driven
nucleus-electron plasma in an erupting flux loop, in which the leptons are
predominantly $e^-$, all of which are accelerated in the same direction.
\vfil
\eject
\centerline{SOURCES AND MECHANISMS}
\bigskip
The release of magnetic energy by reconnection and the release of elastic
energy in starquakes have been considered as possible mechanisms of
GRB at Galactic distances since their discovery$^{3,4,6,13}$.
They are plausible qualitative explanations of much of the phenomenology,
including the observed zoo of diverse GRB shapes and durations.  The
suggestion that much of the energy release and radiation comes from a thin
sheet may also be explicable if the immediate mechanism is magnetic
reconnection.  The gross energetics may be consistent with GRB at 100
Kpc distance$^{4,11,13}$.

It is also possible to consider$^{10}$ hybrid models, in which the eruption
of current loops requires crust-breaking.  The chief difficulty faced by
magnetic reconnection is to release energy suddenly in regions of low
optical depth.  Dissipation requires resistivities between those of the
interior (large) and vacuum (0); turbulent plasmas are plausible.
The details of both physics and field geometry are likely to be complex.
\bigskip
\centerline{A CRAZY IDEA}
\bigskip
Comets in an inner Oort cloud may occasionally collide with velocities $\sim
1$ km/sec.  It is conceivable, if implausible, that such a collision of cold
masses of dirty ice might produce gamma-rays by electrostatic processes.
Deformation of piezoelectric components, heating of pyroelectric components
or frictional charging, as the comets splatter in a subsonic collision, might
lead to multi-MeV potentials.  Pure ice is believed not to be piezoelectric
or pyroelectric, but this has been controversial$^{14}$, and dirty ice or
comets could be more complex.  Ice has been reported$^{14}$ to be
triboluminescent, indicating the production of multi-eV potentials.  To produce
gamma-rays requires that MeV potentials accelerate electrons through the
vacuum.  This is not unprecedented; pyroelectrics accelerate electrons to
sufficient energy to produce X-rays$^{15}$.  Complex intensity histories might
be attributed to complex collision geometries.  Efficiency requires that no
current flow through solids, where electrons would thermalize; this is
consistent with the high resistivity and dielectric strength of cold ice.

Although the microscopic physics of these electrostatic processes is poorly
understood, it is possible to examine the energetics.  Consider a mass $M$
of comets, each of size $a$ and mass $\rho a^3$ ($\rho \approx 1$ gm/cm$^3$),
symmetrically filling a sphere of radius $R$ centered on the Sun.  There
are $N = M/\rho a^3$ comets, with number density $n \sim M / \rho a^3 R^3$.
A collision cross-section $a^2$ leads to a mean collision time $t_c \sim a
\rho R^{7/2} / M (GM_\odot)^{1/2}$.  The total kinetic energy is $E \sim
GM_\odot M / 2R$, and the mean power density at Earth is
$$F \sim {E \epsilon \over 4 \pi R^2 t_c} \sim {GM_\odot M \epsilon \over
8 \pi R^3 t_c} \sim {(GM_\odot)^{3/2} M^2 \epsilon \over 8 \pi a \rho
R^{13/2}}, \eqno(11)$$
where $\epsilon$ is the efficiency of converting kinetic energy to
gamma-rays.  The collision rate is
$${\dot N} \sim {M^2 \sqrt{GM_\odot} \over \rho^2 a^4 R^{7/2}}.
\eqno(12)$$
The observed$^{16}$ $F \sim 3 \times 10^{-12}$ erg/cm$^2$ sec, and for
the BATSE ${\dot N} \sim 10^{-5}$ sec$^{-1}$.  Defining $M_{-3}
\equiv M/10^{-3}M_\odot$ and $\epsilon_{-3} \equiv \epsilon/10^{-3}$ and
combining (11) and (12) yields
$$R \sim {{\dot N}^{2/45} (GM_\odot)^{11/45} M^{4/15} \over \rho^{4/45}}
\left({\epsilon \over 8 \pi F}\right)^{8/45} \sim 3.2 \times 10^{15}
M_{-3}^{4/15} \epsilon_{-3}^{8/45}\ {\rm cm}, \eqno(13)$$
and
$$a \sim {M^{4/15} \over (GM_\odot)^{4/45} \rho^{19/45} {\dot N}^{13/45}}
\left({8 \pi F \over \epsilon}\right)^{7/45} \sim 1.2 \times 10^6
M_{-3}^{4/15} \epsilon_{-3}^{-7/45}\ {\rm cm}. \eqno(14)$$
The collision time is
$$t_c \sim {\rho^{4/15} M^{1/5} (GM_\odot)^{4/15} \over {\dot N}^{2/15}}
\left({\epsilon \over 8 \pi F}\right)^{7/15} \sim 1.0 \times 10^{17}
M_{-3}^{1/5} \epsilon_{-3}^{7/15}\ {\rm sec}, \eqno(15)$$
consistent with the age of the Solar System of $1.5 \times 10^{17}$ sec if
$\epsilon > 10^{-3}$.

This model makes a number of predictions.  It cannot explain any repeating
source.  The distance scale (13) implies a typical parallax of 1 AU/$R
\approx 15^\prime$ if $\epsilon \sim 10^{-3}$.  This may be tested by
overdetermined GRB position measurements, and the model may be excludable
by extant data.  The distribution of GRB on the sky should show an annually
time-periodic dipole moment of $O(1{\rm\ AU}/R)$, with its peak towards the
Sun.  There will be no cyclotron lines, and any annihilation lines will have
no redshift.  Atomic X-ray lines of abundant species may be present.  There
may be a simultaneous visible flash, of unpredictable intensity, resulting
from scintillation by impacting energetic electrons.  The predicted apparent
visual magnitude of an $a = 12$ km comet at a distance $R$ is about 32, but
a comet disrupted by a collision into a spray of fragments could be many
magnitudes brighter, depending on fragment size.  Such a spray would disperse
at an angular rate $\sim 1^{\prime\prime}$/day.  Dust could also be an
effective scatterer of sunlight.  A fraction, perhaps large,
of the kinetic energy of collision ($\epsilon^{-1}$ times the gamma-ray
energy) would be thermalized and reradiated in the near-infrared; the time
scale depends on the fragment size and the thermal conductivity, and is
incalculable.  The hypothesis of gamma-ray production could itself be tested
in laboratory collisions, if the nature of cometary material were well
enough known.

I thank I. A. Smith for discussions.
\bigskip
\centerline{REFERENCES}
\bigskip
\item{1.} C. A. Meegan, G. J. Fishman, R. B. Wilson, W. S. Paciesas, G. N.
Pendleton, J. M. Horack, M. N. Brock and C. Kouveliotou, {\it Nature} {\bf
355}, 143 (1992).
\item{2.} G. J. Fishman, these proceedings.
\item{3.} R. Ramaty, R. E. Lingenfelter and R. W. Bussard, {\it Ap. Sp.
Sci.} {\bf 75}, 193 (1981).
\item{4.} J. I. Katz, {\it Ap. J.} {\bf 260}, 371 (1982).
\item{5.} T. L. Cline, {\it Comments on Ap.} {\bf 9}, 13 (1980).
\item{6.} M. A. Ruderman, {\it Ann. N. Y. Acad. Sci.} {\bf 262}, 164 (1975).
\item{7.} T. L. Cline, in {\it High Energy Transients in Astrophysics}, ed.
S. E. Woosley (AIP, N. Y., 1984), p. 333.
\item{8.} M. C. Jennings, in {\it High Energy Transients in Astrophysics},
ed. S. E. Woosley (AIP, N. Y., 1984), p. 412.
\item{9.} J. I. Katz, {\it High Energy Astrophysics} (Addison-Wesley, Menlo
Park, 1987), p. 284.
\item{10.} B. J. Carrigan and J. I. Katz, {\it Ap. J.} {\bf 399}, 100 (1992).
\item{11.} J. I. Katz, {\it Ap. Sp. Sci.} in press (1992).
\item{12.} M. C. Weiskopf, P. G. Sutherland, J. I. Katz and C. R. Canizares,
{\it Ap. J. (Lett.)} {\bf 223}, L17 (1978).
\item{13.} O. Blaes, R. Blandford, P. Goldreich and P. Madau, {\it Ap. J.}
{\bf 343}, 839 (1989).
\item{14.} P. V. Hobbs, {\it Ice Physics} (Clarendon Press, Oxford, 1974).
\item{15.} J. D. Brownridge, {\it Nature} {\bf 358}, 287 (1992).
\item{16.} M. C. Jennings and R. S. White, {\it Ap. J.} {\bf 238}, 110
(1980).
\vfil
\eject
\bye
\end